# Multivariate scaling of proton and ion energies, divergence, and charge states in Target Normal Sheath Acceleration

Vasiliki E. Alexopoulou [1, *] (https://orcid.org/0000-0003-4068-9810)

[1]School of Mechanical Engineering, Section of Manufacturing Technology, National Technical University of Athens, Heroon Polytechniou 9, 15780 Athens, Greece

**ABSTRACT**. The interaction of an intense laser pulse with a solid target produces energetic proton and ion beams through the Target Normal Sheath Acceleration (TNSA) mechanism. Such beams are under active investigation for applications in proton beam therapy, materials modification, and nuclear and high-energy-density physics. Despite extensive experimental and theoretical effort, predictive correlations between laser and target parameters and the resulting ion-beam properties remain an open research question, owing to the intrinsically multiphysics and strongly coupled nature of laser–plasma interactions. Here, we employ our unified multiphysics model that reproduces laser–solid interaction dynamics with accuracy exceeding 95% over a broad range of short- and ultrashort-pulse conditions. Using this model, we derive statistically validated scaling laws and probability maps that correlate proton, carbon, and oxygen ion cutoff energies, beam divergences, and ionization states to a wide set of laser and target parameters, including pulse duration, laser power, laser beam spot, target thickness, prepulse–main pulse interval, contrast, laser wavelength, and polarization. Continuous beam properties (cutoff energies and beam divergences) are described using multivariate regression with cross-validation, while discrete ionization states are analyzed using classification and regression tree (CART) methods, enabling nonlinear and threshold-dependent behavior to be captured. The resulting scaling relations, contour maps, and box plots elucidate the coupled roles of laser pulse, and target geometry in governing TNSA ion acceleration and charge-state formation. These results provide a predictive and physically interpretable framework for understanding and optimizing laser-driven ion sources across a wide parameter space.

## I. INTRODUCTION

When a metallic target with a thickness ranging from micrometers to millimeters is irradiated by an intense laser pulse (intensity > $10^{18}$ W cm$^{-2}$), electrons within the material are rapidly accelerated to relativistic velocities through a mechanism called Target Normal Sheath Acceleration (TNSA) [1]. Their escape from the irradiated surface generates a strong electrostatic field − commonly referred to as the target normal sheath field. Within this field, protons and heavier ions such as carbon and oxygen, typically originating from surface contaminants or native oxide layers, are subsequently accelerated, producing high-energy, short-duration ion beams. These laser-driven ion beams are of significant interest for applications including advanced cancer proton therapy, ion implantation, laboratory astrophysics, and nuclear science [2,3].

Although the TNSA process unfolds over only a few nanoseconds, it is intrinsically complex because it consists of multiple coupled physical mechanisms occurring sequentially and sometimes simultaneously. Indicatively, these include material ablation and preplasma formation, laser irradiation absorption and electron acceleration, electron transport and scattering inside the target, sheath field formation at the rear surface, and finally ion extraction and acceleration within this rapidly evolving electrostatic field [4]. Due to this multistage nonlinear chain of events, establishing direct relationships between the initial laser and target parameters and the properties of the resulting proton and ion beams remains a significant scientific challenge. Achieving this correlation is essential for gaining a deeper understanding of the TNSA mechanism itself and for enabling predictive optimization of experimental design and ion source performance.

Recent studies have advanced experimental scaling laws for laser-driven proton acceleration, but important limitations remain. Zimmer et al. [5] analyze proton data from 22 published experiments and derive an empirical scaling law for the proton cutoff energy as a function of laser pulse duration, energy, spot size, and target thickness. Their model is tested against the DRACO laser and used to predict performance for ELI-NP, but its accuracy relies on calibration with additional experiments from the same laser system, making it effectively a laser-dependent scaling law rather than a universally predictive model. Simpson et al. [6] perform experiments in the sub-picosecond to multi-picosecond regime using the TITAN laser. They show that hot-electron temperatures − and therefore proton energies − greatly exceed previous model predictions. However, the resulting scaling law is tied to the specific experimental setup with a narrow range of pulse durations and energies, and cannot be assumed to apply to other laser systems. Keppler et al. [7] base their scaling on experiments at the POLARIS laser, a single facility with fixed laser parameters. They demonstrate that proton energy depends on both laser energy and the preplasma scale length, and derive corresponding scaling laws. Again, these laws are strongly system-dependent and cannot be universally applied. Taken together, these studies highlight meaningful progress but also illustrate the following limitations of experimentally derived scaling laws:

• Most scaling laws are based on a relatively small number of experiments ($\approx 10 - 30$) due to their high cost, limiting statistical robustness.

• Many datasets originate from a single laser facility, so the resulting scaling law is intrinsically tied to specific laser architectures, contrast properties, focusing geometries, and diagnostics.

• Even scaling laws based on multi-facility datasets require system-specific tuning before application to a new laser, since they do not account for key experimental conditions such as preplasma effects, making current formulations descriptive rather than universal.

• The majority of investigations focus mainly on proton cutoff energy, with limited experimental analysis of other key parameters such as beam divergence or heavier ion species (e.g., carbon and oxygen).

Modeling of TNSA has advanced through several approaches, each providing insight into specific aspects of the mechanism, while still facing inherent limitations. Theoretical models [8,9] offer a simplified, physically transparent description of sheath field formation and ion acceleration, often allowing for analytical predictions of maximum energies. Particle-in-cell (PIC) simulations [10–12] provide a more detailed treatment of laser–plasma interactions, capturing the dynamics of electrons and ions in the sheath field. Hydrodynamic codes [13] are employed to simulate the formation and expansion of preplasma induced by laser prepulses and pedestal effects on the solid target. Monte Carlo methods [14,15] are used to model the transport and interactions of energetic electrons within the target, including scattering, energy loss, and secondary emission processes. Despite this methodological diversity, three recurring limitations constrain the development of broadly predictive scaling laws based on current modeling approaches:

• Partial coverage of the full physics chain. Each model typically focuses on a single stage of TNSA − preplasma evolution, electron transport, or sheath acceleration − while treating other stages empirically or via imposed boundary conditions. This staged approach can miss important feedback between stages, potentially introducing large systematic errors when linking initial laser parameters to final ion beam characteristics.

• Limited parameter sampling due to computational cost. High-fidelity approaches, such as PIC and hydrodynamic simulations, are resource-intensive, so studies generally explore only a small number of parameter combinations ($\approx 10 - 20$). Sparse sampling reduces statistical robustness and limits the reliability and universality of derived scaling laws.

• Narrow observables. Similar to the experimental literature, most modeling efforts emphasize proton cutoff energy, with limited systematic analysis of other beam metrics, such as divergence, or scaling for heavier ions like carbon and oxygen.

Overall, while existing modeling approaches provide valuable insights into the TNSA mechanism and inform experimental interpretation, their limitations highlight the need for integrated frameworks, coupled with extensive experimental validation to develop broadly predictive scaling laws. In our previous work [16], we developed an analytical multiphysics simulation framework that self-consistently couples heat transfer, hydrodynamics, and electromagnetism, capturing all key physical processes involved in laser-driven particle acceleration via TNSA. Each physical process is treated using its appropriate governing equations, while ensuring that all models are coupled through accurate, self-consistent initial conditions derived from one another. As a result, our framework achieves excellent agreement with experimental data, with predictive accuracy exceeding 95% across multiple major HED laser facilities (TITAN, TPW, OMEGA EP, ORION). This framework possesses several features that make it particularly suitable for developing universal scaling laws:

• Self-consistent multiphysics approach: No empirical inputs are required, allowing accurate predictions across the full range of parameter combinations and supporting highly predictive scaling laws.

• Cross-facility validation: Excellent agreement with experiments from multiple distinct laser systems ensures robustness and broad applicability of the resulting scaling relations.

• Analytical efficiency: The framework can perform up to 1,700 runs with different laser–target parameter combinations, enabling comprehensive coverage of current laser capabilities and achieving strong statistical robustness.

In this paper, we leverage this model to systematically investigate the correlations between all major laser and target parameters − target thickness, laser power, laser beam spot, pulse duration, laser wavelength, polarization, contrast, and prepulse-main pulse interval − and key proton and ion (C, O) beam characteristics, including cutoff energies, divergences, and ionization states. We employ advanced statistical tools to quantify the significance of each input parameter on the final particle characteristics and then apply regression and classification models to derive universal predictive scaling laws and scaling maps.

## II. METHODS

### A. Overview of our analytical multiphysics model

Here, we employ our analytical multiphysics simulation framework previously developed and detailed in [16]. The model self-consistently couples heat transfer, hydrodynamics, and electromagnetic field evolution to simulate laser–solid interactions leading to Target Normal Sheath Acceleration (TNSA). The framework is designed to capture the full physics chain − from initial laser energy deposition to the emergence of ion beams − without reliance on empirical scaling inputs. Only essential methodological details are summarized here; the full formulation and validation appear in [16].

From a physical perspective, TNSA triggered by ultrashort HED laser irradiation of a metallic target (e.g., Au) proceeds through three sequential stages:

1. Thermalization and ablation (Fig. 1(a)). The target surface is heated by short and ultrashort laser prepulses or pedestals. Ultrafast energy deposition leads to an electron–lattice non-equilibrium thermalization and explosive material removal.

2. Preplasma ionization and expansion (Fig. 1(b)). The ablated material is ionized, forming a preplasma that expands away from the surface and defines the preplasma scale length.

3. Main pulse interaction and particle acceleration (Fig. 1(c)). The main HED laser pulse interacts with the formed preplasma, driving the generation and propagation of relativistic hot electrons and subsequently accelerating protons, carbon ions, and oxygen ions originating from surface contaminants or oxide layers.

Correspondingly, each stage is modeled using a dedicated module:

1. Heat transfer module (Fig. 1(d)). Laser and target parameters initialize the heat-transfer calculations, yielding electron and lattice temperature evolution. To accurately describe the non-equilibrium character between electrons and the lattice, our heat transfer module is based on the two-temperature model (TTM). The outputs of this module define the initial state for the hydrodynamics stage.

2. Hydrodynamic module (Fig. 1(e)). This module resolves explosive material removal, ionization, and preplasma expansion, providing estimates of preplasma scale length, temperature, and ionization level. This module is based on Euler equations coupled to SESAME equation-of-state data [17]. The outputs from this module are then introduced to the electromagnetic stage.

3. Electromagnetic module (Fig. 1(f)). Electron dynamics − including superponderomotive acceleration, scattering, reflux, and radiative losses − are modeled to predict the resulting sheath field without external assumptions. Proton and ion acceleration occurs within this self-consistent sheath field.

The framework has been benchmarked against multiple major high-energy-density laser systems (TITAN, TPW, OMEGA EP, ORION), achieving predictive agreement exceeding 95%. This validated, non-empirical formulation makes the model suitable for systematic parameter-space exploration and extraction of statistically robust scaling laws.

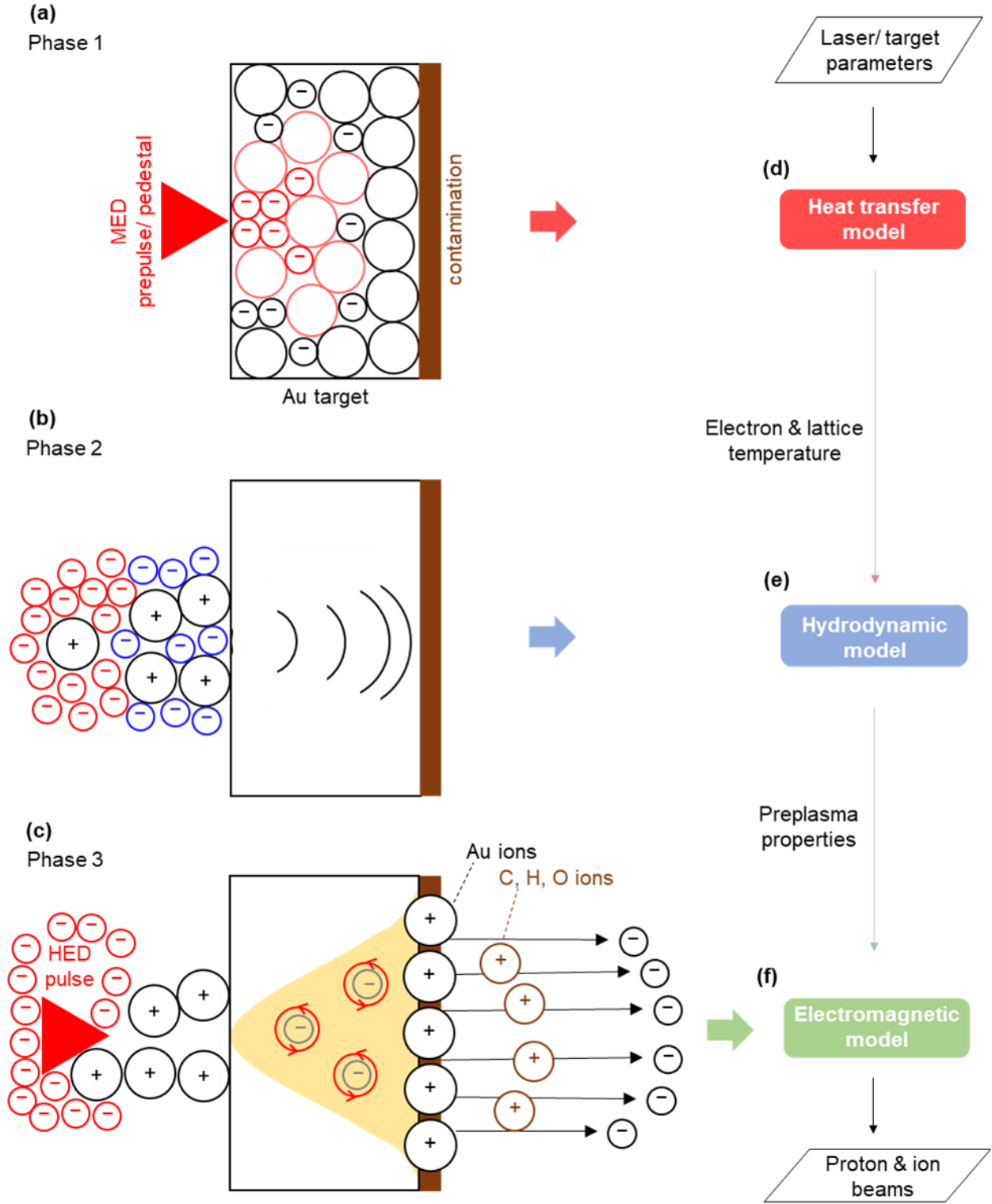

FIG 1. TNSA stages and multiphysics model: Laser prepulses induce nonequilibrium heating and ablation (**a**), described by the heat-transfer module (**d**), followed by preplasma expansion (**b**), simulated by the hydrodynamic module (**e**), and generation of a rear-surface sheath field (**c**) that accelerates protons and heavier ions (C, O), modeled by the electromagnetic module (**f**). Together, the coupled modules capture the full TNSA process.

## B. Ion cutoff energy, divergence, and charge state, estimation

In this work, ion beam properties – namely cutoff energy, most-likely charge state, and divergence – are computed using an analytical TNSA formulation. The approach is based on Mora's self-similar expansion model and is supplemented with an over-the-barrier ionization (OBI) criterion for charge-state determination and a transverse-kick approximation for estimating beam divergence. This analytical treatment is applied separately to protons, carbon ions, and oxygen ions, using sheath parameters obtained from the multiphysics framework.

### 1. Notation and species definition

All subsequent equations are expressed in species-indexed form unless otherwise stated. A generic ion species is indexed as $i$, while explicit cases refer to protons ($p$), carbon ($c$), and oxygen ($o$). The corresponding masses and atomic masses are:

- $m_p, m_c, m_o$

- $Z_p = 1, Z_c = 6, Z_o = 8$

This notation applies consistently throughout the analytical formulation of energy, charge state, and divergence.

### 2. Sheath and plasma parameters

The parameters computed by the multiphysics framework – namely the rear-side hot-electron temperature ($T_e$), electron density ($n_e$), sheath field evolution, and electron divergence – serve as inputs to the analytical ion acceleration model summarized below. This step enables the extraction of ion-beam observables required for scaling law analysis, including cutoff energy, most-likely charge state, and beam divergence.

First, the rear-side hot-electron temperature $T_e$ and rear electron density $n_e$ are used to evaluate the species ion plasma frequency and ion sound speed, respectively [18]:

$$\omega_{pi,i} = \sqrt{\frac{Z_i e^2 n_e}{\varepsilon_0 m_i}} \quad (1)$$

$$c_{s,i} = \sqrt{\frac{Z_i k_B T_e}{m_i}} \quad (2)$$

from which, the characteristic expansion time is set to $\tau_{\exp,i} = 6/\omega_{pi,i}$.

### 3. Acceleration and normalized time

Each species can achieve full acceleration in time estimated by [19]:

$$\tau_{acc,i} = \sqrt{\tau_{pulse}^2 + \tau_{exp,i}^2 + \left(\frac{r_{beam}}{v_d}\right)^2} \quad (3)$$

where $\tau_{pulse}$ is the laser pulse duration, $r_{beam}$ is the laser beam spot radius and $v_d$ is the electron drift velocity.

The normalized time parameter used in Mora's model is given by:

$$t_{p,i} = \frac{\omega_{pi,i}\, \tau_{acc,i}}{2e} \quad (4)$$

### 4. Ion cutoff energy

For protons, our model utilizes the Mora expression for the TNSA cutoff energy [8]:

$$E_{\max,p} = 2\, k_B T_e \left[\ln\left(t_{p,p} + \sqrt{1 + t_{p,p}^2}\right)\right]^2 \quad (5)$$

For heavier ions our model uses the quasi-static model:

$$E_{\max,i} = Z_i^* E_{\max,p} \quad (6)$$

where $Z_i^*$ is the most-likely charge state.

### 5. Sheath field strength, ion front velocity and position

The Debye length and the sheath field strength are computed, respectively, by [20]:

$$\lambda_D = \sqrt{\frac{\varepsilon_0 k_B T_e}{e^2 n_e}} \quad (7)$$

$$E_{front,i} = \sqrt{\frac{2 n_e k_B T_e}{e \varepsilon_0 (1 + t_{p,i}^2)}} \quad (8)$$

The expansion ion front velocity and position follow the standard self-similar expressions, respectively [8]:

$$v_{front,i} = 2 c_{s,i} \ln\left(t_{p,i} + \sqrt{1 + t_{p,i}^2}\right) \quad (9)$$

$$z_{front,i} = 2\sqrt{2e}\,\lambda_D \cdot$$
$$\cdot \left[ t_{p,i} \frac{v_{front,i}}{2c_{s,i}} - \sqrt{1 + t_{p,i}^2} + 1 \right] \quad (10)$$

### 6. Over-the-barrier ionization (OBI) criterion

Carbon and oxygen charge states are determined by an over-the-barrier ionization (OBI) threshold [21] applied to the computed quasi-static sheath field $E_{front,i}$. For each successive ionization potential $I_p^{(k)}$ the corresponding OBI threshold field $F_{OBI}$ is computed, and the highest ionization stage with $E_{front,i} \geq F_{OBI}$ is returned, from which we obtain the most-likely charge state $Z_i^*$.

### 7. Beam divergence (half-angle) computation

Our model uses a simple transverse-kick model [22] to estimate transverse acceleration as a sheath transverse component proportional to the axial field:

$$E_{r,i} = \kappa\, E_{front,i} \quad (11)$$

where the coefficient $\kappa$ depends on the electron beam broadening angle $\theta_e$ [16] and it is estimated by:

$$\kappa = \sin(\theta_e) \quad (12)$$

The transverse velocity gained is integrated over the effective acceleration time $\tau_{eff,i}$:

$$v_r = \frac{Z_i e E_r}{m_i} \tau_{eff,i} \quad (13)$$

For protons, $\tau_{eff,p} = \tau_{acc,p}$. For heavy ions the effective acceleration time is limited by both their ability to reach the front and the duration of the sheath field:

$$\tau_{eff,i} = min\left(\tau_{acc,p}, \frac{z_{front,i}}{v_{front,i}}\right) \quad (14)$$

The longitudinal representative velocity is taken as $v_z = v_{front,i}$ and the output half-angle is:

$$\theta_{1/2,i} = \arctan\left(\frac{v_r}{v_z}\right) \quad (15)$$

## C. Design of experiments and parameter space conditioning

### 1. Factor definition and parameter space

To explore the multidimensional relationship between laser–target parameters and the resulting proton and ion beam characteristics, a structured Design of Experiments (DOE) approach is implemented. Eight experimentally relevant input factors are selected:

- target thickness,
- laser power,
- laser beam spot,
- pulse duration,
- laser wavelength,
- polarization,
- contrast, and
- prepulse–main pulse interval

For each factor, discrete levels spanning the full operational ranges of current HED laser systems are assigned (Table I).

### 2. Candidate matrix generation and subsampling

An initial full-factorial candidate set [22] is generated – an established tool for structured sampling in multidimensional design spaces. A full factorial at these level counts would require several hundred thousand simulations, therefore a uniform random subsampling strategy, called random subsampling without replacement [23], is applied to extract 1,700 unique and non-repeating parameter combinations. This approach preserves the factorial structure at the coded level while ensuring computational tractability and avoiding bias toward any region of the design space. The resulting DOE matrix is then randomized to eliminate run-order correlation effects. Finally, each coded level is subsequently mapped to its corresponding physical value.

### 3. Feasibility filtering based on irradiance

Not all factor levels are physically compatible with one another. In particular, certain combinations of laser power and beam spot size (e.g., very high power with very small focal spots, or the opposite) produce irradiances either unrealistically high or too low to generate meaningful acceleration. To address this, an irradiance feasibility check [24] is applied, where irradiance is calculated as:

$$I = \frac{P_{laser}}{\pi \left(\frac{d_{beam}}{2}\right)^2} \quad (16)$$

where $P_{laser}$ is the laser power and $d_{beam}$ is the laser beam spot diameter.

Only parameter sets producing irradiances within the experimentally relevant range of $10^{18} - 10^{21}$ W/cm$^2$ are retained [25]. This ensures that the DOE remains representative of realistic laser–solid interaction conditions associated with TNSA. This first physical filtering reduces the initial DOE to 1,002 feasible parameter sets, ensuring that subsequent analysis represents realistic TNSA-relevant conditions. All 1,002 feasible parameter sets pass forward as valid inputs to our multiphysics model.

Table I. Design of Experiments: Factors, levels, and justifications.

| Factor | Levels | Justification |
|---|---|---|
| Target thickness (μm) | 1 – 10 – 100 – 1000 | Covers ultrathin to bulk targets, allowing comparison of short-scale TNSA behavior and thicker-target transport effects. |
| Laser power (TW) | 20 – 250 – 500 – 800 – 1000 | Spans moderate to near-PW systems, enabling study of how acceleration scales with available energy. |
| Laser beam spot (μm) | 2 – 10 – 20 | Beam diameter size directly sets intensity and affects sheath formation and beam divergence. |
| Pulse duration (fs) | 10 – 100 – 1000 – 10000 | Includes ultrashort to short pulses, capturing the full range of available HED pulses. |
| Laser wavelength (nm) | 815 – 1054 | Represents the most common high-power laser systems (Ti:Sa and Nd:YAG/OPCPA). |
| Polarization | Linear − Circular | Polarization influences electron heating efficiency and sheath symmetry. |
| Contrast | $10^{-8}$ – $10^{-7}$ – $10^{-6}$ – $10^{-5}$ | Determines preplasma formation, which strongly affects the sheath structure and acceleration efficiency. |
| Prepulse-main pulse interval (ns) | 0.03 – 0.3 – 3 | Controls preplasma scale length, affecting ionization and early hydrodynamic conditions. |

### *4. Post-simulation output filtering based on proton cutoff energy*

A second feasibility filter [24] is applied to address output-based inconsistencies. Certain combinations of contrast, prepulse–main pulse interval, pulse duration, and target thickness can produce unrealistic proton and ion energies. For example, a low contrast (typical for prepulses), long prepulse–main pulse interval (typical for pedestals), long pulse duration, and ultrathin target may generate excessively expanded preplasmas, resulting in artificially high cutoff energies not observed experimentally. Conversely, the opposite extreme − high contrast, short interval, long pulse duration, and thick targets − can strongly attenuate hot electrons within the material, preventing rear-surface sheath formation and thus yielding negligible proton and ion acceleration.

Because such behavior cannot be predicted a priori, this second filtering step is conducted after running the multiphysics simulations. Specifically, cases producing either no proton and ion acceleration or non-physical proton cutoff energies exceeding 85 MeV are excluded [26]. This upper limit reflects the application-driven performance window and prevents extreme outliers from biasing subsequent statistical modeling and regression analyses. After this post-processing stage, 517 valid cases remain.

Of the initial 1,700 DOE entries, 1,002 passed the first irradiance-based filter and only 517 passed the second energy-based filter. This substantial reduction is expected due to the high dimensionality of the parameter space: many factor combinations correspond to physically infeasible laser–target conditions, such as irradiances outside operational limits or extreme prepulse/pulse/target scenarios that either over-accelerate ions or suppress sheath formation entirely.

Filtering ensures that only physically meaningful and experimentally relevant cases are retained while preserving the broad coverage and unbiased structure of the original DOE. The final dataset, consisting exclusively of physically plausible and operationally meaningful cases, is exported in spreadsheet format and used for regression, sensitivity studies, and extraction of scaling laws.

### *5. Non-orthogonality of the DOE and implications for statistical analysis*

It is important to highlight that the final DOE is non-orthogonal. In an orthogonal DOE, each factor varies independently from the others, making it straightforward to separate and quantify individual effects using classical ANOVA Type I-based methods. In our case, however, several factors are physically linked (for example, laser power and beam spot size together define irradiance), meaning they cannot vary freely or independently across all level combinations. Additional filtering steps based on physical feasibility further increase these correlations. As a result, some predictors influence the response jointly rather than in isolation, and their effects cannot be interpreted using orthogonal assumptions. For this reason, alternative statistical tools (based on ANOVA Type III) that account for correlated variables − such as regression modelling with backward term elimination and cross-validation − are required [27]. These methods allow meaningful extraction of scaling behavior while maintaining consistency with the underlying physics and constraints of high-intensity laser–target interactions.

### D. Statistical analysis and scaling law extraction for continuous response variables

To quantify the influence of laser–target parameters on the resulting proton and ion beam properties and to derive scaling laws, a statistical regression framework is implemented. The analysis focuses on six continuous response variables: proton, carbon, and oxygen cutoff energies and half-angle divergences.

### *1. Model formulation and data transformation*

Scaling relations for laser-driven ion acceleration are widely known to follow power-law behavior. Since regression [28] is carried out in linear form, both the predictors and response variables are transformed using the natural logarithm. This allows exponential relationships of the form:

$$Y \propto \prod_i X_i^{\alpha_i} \qquad (17)$$

to be expressed as a linear additive model:

$$\ln(Y) = \beta_0 + \sum_{i=1}^{n} \beta_i \ln(X_i) + \epsilon \qquad (18)$$

where $\epsilon$ represents the residual error accounting for variability not explained by the predictors. Following model fitting, the equations are reverted to their physical form by exponentiation.

### *2. Parameter significance assessment*

To determine which physical parameters most strongly influence the model outputs and to remove statistically redundant terms, a backward elimination strategy [29] is applied. The full regression model is initially constructed with all available predictor variables. Then, at each iteration, the predictor with the highest p-value (i.e., the least statistically significant term) is removed, provided its p-value exceeded the significance threshold $\alpha = 0.05$. After each elimination step, the regression is recalculated to update the coefficient estimates and significance levels. The procedure continues until all remaining parameters meet the significance criterion. The p-values used for removal are obtained from t-tests applied to each regression coefficient, testing the null hypothesis that the true parameter effect is zero.

This method is particularly suitable for the present dataset because the DOE is non-orthogonal [30]. During the design process, several parameter combinations are excluded due to physical constraints, resulting in correlated predictors. Unlike orthogonal designs, where parameter effects are estimated independently, non-orthogonal designs require that each parameter's contribution is evaluated while accounting for interdependencies with others. Backward elimination inherently supports this, as the statistical significance of each predictor is reassessed after each model reduction step, ensuring that correlated terms are evaluated in context rather than in isolation.

### 3. Assessing predictor independence using Variance Inflation Factor (VIF)

To further assess whether the regression model reliably separates the effect of each predictor, Variance Inflation Factor (VIF) values are calculated [28]. VIF quantifies the degree to which the variance of a regression coefficient is inflated due to multicollinearity with other predictors. A VIF value of 1 indicates no correlation, while values above approximately 5 suggest strong multicollinearity and reduced reliability of coefficient estimates. In this analysis, VIF values are extracted directly from the regression diagnostics at each model iteration. High VIF values would indicate that the backward elimination method may not reliably separate the individual effects of correlated predictors. In such cases, more advanced techniques, such as Principal Component Regression (PCR) or Partial Least Squares (PLS), would be required to obtain robust and interpretable parameter estimates.

By combining backward elimination, t-statistical testing, and VIF screening, the final regression model retains only statistically significant, non-redundant, and minimally correlated predictors. This ensures that the resulting model is interpretable, avoids overfitting, and provides a robust basis for physically meaningful scaling law extraction.

### 4. Model validation and performance assessment

Model robustness is assessed using 10-fold cross-validation [29]. The dataset is randomly partitioned into ten equally sized subsets. For each validation cycle:

- nine subsets are used for model training, and

- the remaining subset is used for testing.

Validation metrics – including $R^2$, adjusted $R^2$, and 10-fold cross-validated $R^2$ – are estimated to evaluate generalization performance and ensure that the resulting scaling laws are not artifacts of specific training subsets.

### 5. Final scaling law construction

Once the optimal model structure is identified and validated, the final regression equation is exponentiated to obtain closed-form scaling laws. The general form of the resulting expressions is:

$$Y = A \prod_i X_i^{\alpha_i} \qquad (19)$$

where $Y$ is the predicted cutoff energy or divergence, $X_i$ are the physical input parameters (e.g., laser power, pulse duration, contrast, etc.), $\alpha_i$ are the fitted scaling exponents, and $A = e^{\beta_0}$ is the prefactor extracted from the linear model intercept.

These final laws directly provide interpretable quantitative insight into how each physical control parameter contributes to the scaling of proton and ion beam performance.

## E. Statistical analysis and scaling for discrete response variables

The ionization state of carbon and oxygen ions is a categorical (non-continuous) output and therefore cannot be modeled using standard linear regression. Instead, a classification-based statistical approach is applied using Classification and Regression Trees (CART).

### 1. Assessing factor influence

To determine the significance and predictive power of each input parameter, a CART model [31] is constructed using:

- Gini impurity as the splitting criterion, and

- 10-fold cross-validation for performance assessment and overfitting control.

From this analysis, the following diagnostic outputs are extracted:

• Multinomial response probabilities, providing the likelihood of each ionization state within the explored parameter ranges.

• Relative variable importance ranking, indicating which experimental factors contribute most strongly to ion charge-state variation.

• Confusion matrices for both training and validation sets, used to quantify classification accuracy and identify misclassification trends.

*2. Generating usable scaling representations*

Although CART method produces a classification tree, the optimal tree structure contains a large number of decision nodes ($\approx 30 - 100$). While accurate, such a structure is not easily interpretable and lacks direct physical meaning. For publication and practical use, a more readable scaling representation is therefore adopted.

Instead of using the full decision tree, the ionization behavior is presented as a set of probability contour maps and box plots, where the probability of each ionization state is plotted as a function of two input factors at a time. These maps provide:

• A physically intuitive visualization of the charge-state evolution,

• a compact representation of multidimensional behavior, and

• a format suitable for experimental design and prediction.

The selection of factor pairs used in the contour plots is guided by the relative variable importance results, ensuring that the most physically and statistically influential parameters are emphasized.

### III. RESULTS AND DISCUSSION

#### A. Significance analysis and scaling of continuous outputs

*1. Parameter significance and multicollinearity*

The significance of each predictor and potential multicollinearity are first assessed using the backward elimination approach, summarized in a heatmap (Fig. 2) and a bar chart (Fig. 3). The heatmap displays the p-values of all factors for each continuous output (proton, carbon, and oxygen cutoff energies, and half-angle divergences), while the bar chart shows the corresponding VIF values.

From the p-values heatmap (Fig. 2), all factors except polarization are statistically significant ($p < 0.05$) for all outputs. Polarization is not significant for any output and is therefore excluded from the scaling laws. Laser wavelength is marginally significant for cutoff energy-related outputs ($p = 0.043 - 0.048$) and is retained in the scaling laws to account for subtle wavelength-dependent effects. The remaining parameters (target thickness, laser power, laser beam spot, pulse duration, prepulse–main pulse interval, and contrast) all have p-values much smaller than 0.05, confirming their strong influence on the outputs.

Additionally, the VIF bar chart (Fig. 3) indicates that all predictors have VIF values below 1.3, demonstrating low multicollinearity. This validates that backward elimination is appropriate for this dataset and that the resulting regression coefficients are not biased by interdependencies between parameters.

*2. Discussion on multivariate dependence of TNSA scaling*

The fact that most parameters (Fig. 2) exhibit very strong statistical significance highlights their collective physical importance in governing sheath formation, electron heating, and ultimately ion acceleration. This outcome reinforces a broader physical interpretation: TNSA does not depend on a single dominant variable, but instead emerges from the coupled interplay of target thickness, laser power, laser beam spot, pulse duration, prepulse-main pulse interval, and contrast. These parameters simultaneously influence the hot-electron temperature, number density, divergence, and confinement at the rear surface − each of which contributes nonlinearly to the magnitude and evolution of the sheath field.

This finding is also meaningful in the context of existing scaling models. Earlier scaling relations in the literature [5–7] frequently rely on reduced parameter sets − typically using only peak intensity and pulse duration (or occasionally intensity and target thickness) to approximate ion cutoff energy. While such simplified descriptions offer useful intuition, they inherently neglect coupled mechanisms including preplasma expansion, laser absorption efficiency, electron refluxing, and geometric effects, all of which modify the effective sheath formation process. As a result, reduced-parameter scaling can overestimate or underestimate ion energy when applied outside the narrow regime in which they were originally derived.

|  | Cutoff energy | | | Half-angle divergence | | | |
| --- | --- | --- | --- | --- | --- | --- | --- |
|  | Proton | Carbon | Oxygen | Proton | Carbon | Oxygen | |
| | <0.001 | <0.001 | <0.001 | <0.001 | <0.001 | <0.001 | Target thickness |
| | <0.001 | <0.001 | <0.001 | <0.001 | <0.001 | <0.001 | Laser power |
| | <0.001 | <0.001 | <0.001 | <0.001 | <0.001 | <0.001 | Laser beam spot |
| | <0.001 | <0.001 | <0.001 | <0.001 | <0.001 | <0.001 | Pulse duration |
| | <0.001 | <0.001 | <0.001 | <0.001 | <0.001 | <0.001 | Prepulse-main pulse interval |
| | 0.001 | 0.004 | 0.006 | <0.001 | <0.001 | <0.001 | Contrast |
| | 0.043 | 0.048 | 0.043 | <0.001 | <0.001 | <0.001 | Laser wavelength |
| | 0.564 | 0.722 | 0.706 | 0.5 | 0.455 | 0.488 | Polarization |

FIG 2. Heatmap of parameter significance based on p-values: This heatmap presents the p-values obtained from the regression analysis. Green and yellow indicate low to moderate p-values, corresponding to statistically significant effects; parameters in these regions are retained in the scaling laws. In contrast, red highlights non-significant effects, indicating parameters that are excluded from the scaling analysis.

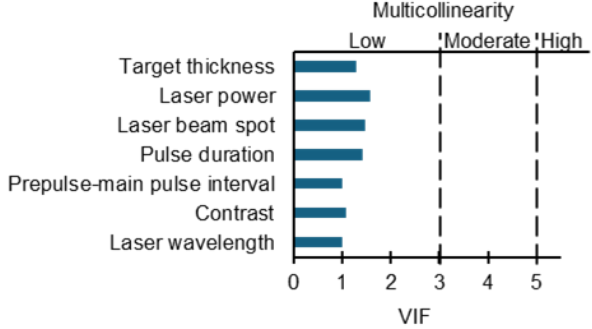

FIG 3. Bar chart of multicollinearity based on VIF values: This bar chart displays the VIF values calculated during the regression analysis using backward elimination. The vertical dashed lines at VIF = 3 and VIF = 5 indicate commonly accepted thresholds for moderate and high multicollinearity, respectively. All parameters exhibit VIF values well below these thresholds, confirming minimal multicollinearity and ensuring that the regression coefficients are robust and interpretable. This validates the suitability of backward elimination for modeling and supports the reliability of the resulting scaling laws.

By contrast, the present analysis demonstrates that predictive capability improves substantially when the complete physical parameter space is considered. The strong statistical weight of these parameters agrees with high-fidelity experimental and numerical studies [3,32], which similarly report that multivariate descriptions are essential for robust and transferable scaling of TNSA ion output. Thus, the results here not only support but quantitatively reinforce the emerging consensus that full-parameter models are necessary to accurately describe and predict ion acceleration in realistic TNSA conditions.

### 3. Discussion on wavelength effects and relevance to TNSA

The statistical analysis (Fig. 2) indicates that laser wavelength has only marginal significance for the outputs considered. This result suggests that, within the operational regime studied here, wavelength plays a measurable but secondary role compared to other parameters such as laser power, laser beam spot, pulse duration, contrast, prepulse-main pulse interval, and target thickness. The relatively weak wavelength dependence can be explained by the underlying physics of electron heating in the TNSA regime.

In principle, wavelength influences key physical quantities such as the normalized vector potential $a_0$, the critical plasma density $n_c \propto \lambda^{-2}$, and the laser skin depth [33]. These factors affect how efficiently the laser couples energy into hot electrons. For example, shorter wavelengths increase the critical density and reduce the skin depth, leading to steeper absorption gradients and potentially more efficient vacuum or Brunel heating. Longer wavelengths, result in lower critical density and deeper penetration, which may slightly modify preplasma formation and electron dynamics [34].

However, in the relativistic intensity regime explored here, the differences induced by the wavelength range

(815 – 1054 nm) are largely masked by dominant processes such as stochastic heating, ponderomotive acceleration, and coupling to plasma waves, all of which scale strongly with intensity rather than wavelength [2]. Moreover, once a preformed preplasma exists, laser energy absorption becomes governed mainly by self-generated fields, and stochastic heating mechanisms − processes that are only weakly sensitive to small changes in wavelength. This observation is consistent with previous experimental and numerical studies, where wavelength effects in TNSA are present but significantly weaker than predicted by earlier theoretical scaling laws derived under simplified assumptions [33]. It is likewise shown that the dependence of cutoff energy on wavelength becomes less pronounced when intensity, contrast, and target conditions push the interaction into the strongly relativistic regime.

Therefore, the marginal wavelength dependence observed here does not indicate a lack of physical relevance, but rather confirms that wavelength sensitivity is suppressed in classical TNSA under realistic experimental conditions. The statistical outcome aligns with current understanding: wavelength plays a measurable but secondary role and is retained in the scaling laws to capture these subtler contributions without dominating the model.

### *4. Discussion on polarization effects and relevance to TNSA*

The statistical analysis (Fig. 2) shows that laser polarization does not significantly affect the continuous outputs (cutoff energies and half-angle divergence) investigated in this work, and therefore it is excluded from the final scaling laws.

This outcome is consistent with the physical understanding of TNSA [35]. In TNSA, protons and ions are primarily accelerated by the quasi-static sheath field established at the rear surface of the target as hot electrons escape. The strength of this sheath field depends predominantly on the total number and temperature of hot electrons rather than on the phase structure or temporal evolution of the driving laser field. As long as sufficient hot electrons are generated, their collective transport and thermalization dominate the subsequent ion acceleration stage.

From a physics perspective [34], laser polarization only weakly influences electron heating in the relativistic regime used here. For linearly polarized light, the electric field direction oscillates, driving electrons into the vacuum and back into the target (vacuum heating). Circularly polarized pulses, by contrast, exert a smoother ponderomotive pressure with reduced oscillatory motion. However, once relativistic intensities are reached and a preplasma forms at the target front, electron heating becomes governed mostly by stochastic processes, self-generated plasma waves, and absorption mechanisms that are comparatively insensitive to the laser polarization state. Therefore, variations in polarization do not significantly modify the resulting sheath field strength or the ion energy distribution for the thickness and intensity regime characteristic of TNSA studied here.

In contrast, while polarization does not influence the TNSA regime investigated here, its role becomes significant in other acceleration mechanisms such as radiation pressure acceleration (RPA) [36,37], which emerges only for ultrathin targets (typically < 100 nm) and requires excellent pulse contrast to prevent premature target expansion. In this regime, circular polarization is necessary to suppress electron oscillations and enable stable momentum transfer from the laser to the target. Similar sensitivity to polarization is also present in relativistic transparency-based regimes [38,39], where polarization influences the onset of volumetric heating and coupling to plasma waves. The absence of such conditions in the present study confirms that the proton and ion acceleration mechanism remains firmly in the TNSA regime, where polarization plays only a minor role. This contextual distinction supports the statistical observation that polarization can be excluded from the scaling laws without loss of physical validity.

Given this mechanistic distinction, the lack of sensitivity to polarization observed in the present dataset is fully consistent with prior detailed studies of TNSA [40,41], where polarization is shown to play a secondary or negligible role compared to other parameters such as target thickness, pulse duration, laser power, contrast, prepulse-main pulse interval, and laser beam spot. Thus, the statistical outcome reinforces existing theoretical and experimental understanding: for classical TNSA under μm- or mm-scale target conditions, polarization is not a governing parameter in determining cutoff energies or ion beam divergence.

### *5. Scaling law derivation and statistical validation*

Using the selected predictors, regression models are fitted for each continuous output. Each scaling law is expressed in exponential form by applying a logarithmic transformation to both input factors and outputs during regression, then exponentiating the resulting equations.

These scaling laws are given below:

- Proton cutoff energy:

$$E_{cutoff}^{proton}[MeV] = A_p d_{t,\mu m}^{-0.36} P_{TW}^{0.78} d_{beam,\mu m}^{-1.29} \tau_{pulse,fs}^{0.58} C^{0.05} \Delta t_{ns}^{0.25} \quad (20)$$

with

$$A_p = \begin{cases} 3.19, for\ \lambda = 815\ nm \\ 3.67, for\ \lambda = 1054\ nm \end{cases} \quad (21)$$

- Carbon ion cutoff energy:

$$E_{cutoff}^{carbon}[MeV] = A_c d_{t,\mu m}^{-0.39} P_{TW}^{0.86} d_{beam,\mu m}^{-1.40} \tau_{pulse,fs}^{0.51} C^{0.04} \Delta t_{ns}^{0.27} \quad (22)$$

with

$$A_c = \begin{cases} 14.79, for\ \lambda = 815\ nm \\ 17.17, for\ \lambda = 1054\ nm \end{cases} \quad (23)$$

- Oxygen ion cutoff energy:

$$E_{cutoff}^{oxygen}[MeV] = A_o d_{t,\mu m}^{-0.44} P_{TW}^{0.95} d_{beam,\mu m}^{-1.48} \tau_{pulse,fs}^{0.48} C^{0.04} \Delta t_{ns}^{0.28} \quad (24)$$

with

$$A_o = \begin{cases} 19.34, for\ \lambda = 815\ nm \\ 22.78, for\ \lambda = 1054\ nm \end{cases} \quad (25)$$

- Proton half-angle divergence:

$$\theta_{half-angle}^{proton}[deg] = B_p d_{t,\mu m}^{0.97} P_{TW}^{-0.90} d_{beam,\mu m}^{1.73} \tau_{pulse,fs}^{-0.49} C^{-0.06} \Delta t_{ns}^{-0.44} \quad (26)$$

with

$$B_p = \begin{cases} 0.29, for\ \lambda = 815\ nm \\ 0.20, for\ \lambda = 1054\ nm \end{cases} \quad (27)$$

- Carbon ion half-angle divergence:

$$\theta_{half-angle}^{carbon}[deg] = B_c d_{t,\mu m}^{0.95} P_{TW}^{-0.89} d_{beam,\mu m}^{1.77} \tau_{pulse,fs}^{-0.45} C^{-0.06} \Delta t_{ns}^{-0.46} \quad (28)$$

with

$$B_c = \begin{cases} 0.06, for\ \lambda = 815\ nm \\ 0.04, for\ \lambda = 1054\ nm \end{cases} \quad (29)$$

- Oxygen ion half-angle divergence:

$$\theta_{half-angle}^{oxygen}[deg] = B_o d_{t,\mu m}^{0.96} P_{TW}^{-0.89} d_{beam,\mu m}^{1.77} \tau_{pulse,fs}^{-0.44} C^{-0.06} \Delta t_{ns}^{-0.46} \quad (30)$$

with

$$B_o = \begin{cases} 0.06, for\ \lambda = 815\ nm \\ 0.04, for\ \lambda = 1054\ nm \end{cases} \quad (31)$$

Here, $d_{t,\mu m}$ is the target thickness in μm, $P_{TW}$ the laser power in TW, $d_{beam,\mu m}$ the laser beam spot diameter in μm, $\tau_{pulse,fs}$ the pulse duration in fs, $C$ the contrast, $\Delta t_{ns}$ the prepulse-main pulse interval in ns, and $\lambda$ the laser wavelength.

To assess scaling models' performance and robustness, three complementary statistical measures are used: $R^2$, adjusted $R^2$, and the 10-fold cross-validated $R^2$ (Fig. 4). The coefficient of determination ($R^2$) quantifies the proportion of variance in the response explained by the model. However, because $R^2$ always increases when additional terms are added − regardless of whether they carry true explanatory value − the adjusted $R^2$ provides a corrected measure that penalizes unnecessary predictors, making it more suitable for evaluating model parsimony. The 10-fold cross-validated $R^2$ further extends this assessment by evaluating predictive accuracy on unseen data through repeated partitioning of the dataset, thereby providing a direct estimate of generalizability.

In this study, these three values are found to be approximately equal across all outputs. This convergence carries important implications:

- The similarity between $R^2$ and adjusted $R^2$ indicates that the included predictors contribute meaningful explanatory power, and that the model is not artificially inflated by redundant variables.

- The agreement between $R^2$ or adjusted $R^2$ and 10-fold cross-validated $R^2$ demonstrates that the regression structure remains stable when exposed to new data, confirming that the model generalizes well rather than fitting noise or dataset-specific fluctuations.

Taken together, this alignment strongly suggests that the developed regression-based scaling laws are statistically robust, free from overfitting, and capable of reliably capturing the underlying physical dependencies governing TNSA ion acceleration.

|  | Cutoff energy | | | Half-angle divergence | | | |
|---|---|---|---|---|---|---|---|
|  | Proton | Carbon | Oxygen | Proton | Carbon | Oxygen | |
|  | 72% | 70% | 70% | 99% | 99% | 99% | $R^2$ |
|  | 71% | 70% | 69% | 98% | 99% | 99% | Adjusted $R^2$ |
|  | 71% | 69% | 69% | 98% | 99% | 99% | 10-fold cross-validated $R^2$ |

FIG 4. Heatmap of scaling laws' performance based on R² statistics: This heatmap compares three goodness-of-fit metrics − R², adjusted R², and 10-fold cross-validated R² − for each ion species and output observable. Green shading indicates very strong predictive performance, while yellow denotes good predictive strength. The close agreement between the three metrics across all outputs demonstrates that the regression models do not suffer from overfitting and generalize well to unseen data. This consistency also confirms that the derived scaling laws capture the underlying physical dependencies governing TNSA acceleration with high reliability.

The performance metrics (Fig. 4) also provide context for interpreting our models' quality across different outputs. For ion cutoff energies, the $R^2$ values fall in the good agreement range (~70%). While these values are lower than those achieved for angular divergence (> 98%), this behavior is fully expected from the physics [35,42]: cutoff energy is more sensitive to micro-scale plasma dynamics, instabilities, and stochastic sheath evolution − processes not fully captured by macroscopic parameters alone. In contrast, beam divergence is governed by more deterministic geometric and temporal characteristics of the accelerating sheath, leading to substantially higher predictability.

Notably, the $R^2$ values achieved here for cutoff energy are comparable to − and in several cases exceed − those reported in earlier regression-based TNSA studies [7,32], which typically relied on reduced parameter sets (e.g., intensity and target thickness alone). However, unlike those simplified models, the present work incorporates a broader multivariate framework spanning laser, target, and prepulse-related parameters across the full experimentally relevant ranges. This means the model is challenged with a substantially more diverse parameter space, increasing the statistical difficulty while offering a far more realistic and generalizable representation of TNSA processes.

An additional distinction is methodological: earlier models were rarely evaluated on unseen data and were often trained and interpreted solely based on fit quality. In contrast, the present scaling laws are evaluated using 10-fold cross-validated $R^2$, which closely matches the original $R^2$ values. This agreement indicates that the models retain predictive accuracy beyond the training dataset and are not the result of overfitting or excessive parameterization.

Taken together, these outcomes demonstrate that even though cutoff energies inherently exhibit higher physical variability than divergence metrics, the proposed multivariate scaling laws achieve strong predictive performance, outperforming or matching existing literature while providing broader applicability and demonstrated robustness to unseen data.

### B. Classification analysis and scaling of categorical outputs

To investigate the dependence of categorical outputs (carbon and oxygen ionization states) on the laser and target parameters, CART models are employed. CART is particularly well-suited for categorical outputs because it can capture nonlinear decision boundaries and complex interactions between predictors without requiring linear relationships or specific distributional assumptions [31]. This capability makes it ideal for modeling discrete charge states, which are highly sensitive to threshold-like behaviors in plasma temperature and electron heating. Three complementary metrics are used to characterize the model behaviour:

• multinomial response summary,
• relative variable importance, and
• confusion matrices for both training and test datasets.

To facilitate physical interpretation and scaling, contour and box plots are constructed to show the probability of each ionization state as a function of laser and target input parameters, providing an intuitive visualization of parameter influences.

#### 1. Multinomial response summary

A multinomial response summary [43] reports the distribution of observations across all discrete categorical outcomes in a multinomial (multi-class) classification problem. It shows how frequently each class occurs in the dataset, providing crucial context for model imbalance, class dominance, and interpretability of classification performance.

The multinomial response table (Table II) for carbon and oxygen reveal clear patterns in the frequency of observed ionization states, reflecting the underlying plasma conditions generated in the TNSA regime.

TABLE II. Multinomial response summary of carbon and oxygen ionization states.

| Carbon ionization state distribution | | | Oxygen ionization state distribution | | |
|---|---|---|---|---|---|
| Charge | Count | % | Charge | Count | % |
| $C^{1+}$ | 20 | 3.9 | $O^{1+}$ | 37 | 7.2 |
| $C^{2+}$ | 38 | 7.4 | $O^{2+}$ | 40 | 7.7 |
| $C^{3+}$ | 39 | 7.5 | $O^{3+}$ | 60 | 11.6 |
| $C^{4+}$ | 379 | 73.3 | $O^{4+}$ | 76 | 14.7 |
| $C^{5+}$ | 41 | 7.9 | $O^{5+}$ | 24 | 4.6 |
| $C^{6+}$ | – | – | $O^{6+}$ | 280 | 54.2 |

### 2. Discussion of the carbon ionization states distribution

The multinomial response shows a strong peak at $C^{4+}$ ($\approx$ 73%), with all other charge states occurring at much lower frequencies, and $C^{6+}$ not observed at all.

This distribution reflects the characteristic electron temperatures produced under classical TNSA conditions. Carbon requires 11.3 eV, 24.4 eV, 47.9 eV, and 64.5 eV for the first four ionization steps, while higher states ($C^{5+}$, $C^{6+}$) require substantially more energy ($\approx$ 392 eV, 490 eV) due to the tightly bound K-shell electrons [44].

Although the hot electron temperatures generated in TNSA can reach up to 85 MeV, the relevant factor for ionization is not only the peak energy but also the electron density at the required threshold energy and the time available for collisional ionization [45]. In practice, most electrons occupy the high keV to low MeV range, making the lower ionization stages ($C^{1+}$–$C^{4+}$) readily accessible. Achieving $C^{5+}$ or $C^{6+}$ requires both a sustained population of very high-energy electrons and sufficient interaction time, conditions that are less consistently met in TNSA than in other acceleration mechanisms such as BOA or RPA [46].

Consequently, $C^{4+}$ emerges as the most probable ionization state (also observed experimentally by other researchers [47]), representing the regime where the electron distribution most effectively overlaps with the carbon ionization thresholds. The absence of $C^{6+}$ is expected, as full K-shell stripping demands electrons of tens of MeV; while such electrons exist in the tail of the distribution, their density and effective interaction time are insufficient to produce significant populations of fully stripped carbon in the majority of laser shots.

### 3. Discussion of oxygen ionization states distribution

Unlike carbon, oxygen exhibits a bimodal-like ionization distribution, with $O^{6+}$ dominating ($\sim$ 54%) and intermediate states ($O^{3+}$ – $O^{4+}$) appearing at moderate frequencies, while $O^{5+}$ remains underpopulated. This pattern reflects the interplay between oxygen's higher nuclear charge, electron-shell structure, and the typical hot-electron spectrum generated in TNSA. The electron populations in the plasma can be described by a multi-peak quasi-Maxwellian distribution, where most electrons cluster around characteristic energies [33]. Electrons at the lower energy spectrum populate the lower charge states ($O^{1+}$ – $O^{4+}$), whereas the higher energy electrons are capable of directly producing $O^{6+}$. The intermediate energy electrons required for $O^{5+}$ are comparatively scarce in this distribution, explaining its lower occurrence. Similar ionization patterns for oxygen under TNSA-relevant conditions have also been reported in previous experimental studies [48].

Higher ionization stages, such as $O^{7+}$ or $O^{8+}$, require stripping K-shell electrons, which demands higher electron energies, largely inaccessible under the classical TNSA conditions considered here. These extreme ionization stages would only appear under more extreme regimes, such as RPA or BOA [49]. Overall, the observed distribution aligns with the "sweet spot" of the electron energy spectrum in TNSA-driven plasmas, where $O^{6+}$ naturally emerges as the most probable charge state.

### 4. Comparison of carbon vs oxygen ionization states distribution

The differing distributions between carbon and oxygen are physically expected. Carbon saturates at $C^{4+}$, because forming $C^{5+}$ and $C^{6+}$ requires K-shell ionization energies that are too high. Oxygen saturates at $O^{6+}$, because the L-shell energies fall within the most populated region of the hot-electron spectrum. Thus, both species exhibit a dominant charge state, determined by the accessible electron-energy band. This alignment confirms that the imbalanced multinomial distribution is physically consistent with known TNSA heating characteristics [46,48,49].

### 5. Relative variable importance

The relative variable importance analysis (Fig. 5) reveals distinct sensitivities of carbon and oxygen ionization states to the laser and target parameters. For both species, pulse duration emerges as the dominant variable (100%), confirming that the energy deposition timescale is the primary factor governing hot-electron generation and, consequently, ionization potential crossing. Target thickness and laser power follow as the next most influential parameters, but with markedly different magnitudes between the two species. Carbon shows strong dependence on both target thickness (90.0%) and laser power (82.9%), whereas oxygen exhibits much weaker sensitivities (53.2% and 31.9%, respectively). This indicates that carbon charge states are primarily controlled by the main pulse interaction and the efficiency of rear surface sheath formation, both of which scale strongly with laser–target coupling and energy delivery.

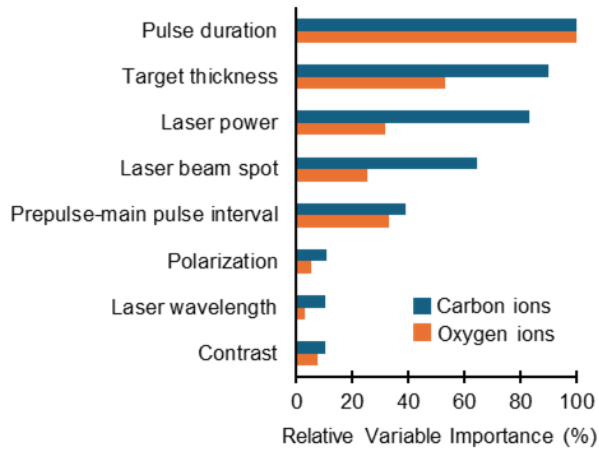

FIG 5. Relative variable importance for predicting carbon and oxygen ionization states: The bar chart presents the normalized relative importance of each predictor in determining the resulting charge states for carbon (blue bars) and oxygen (orange bars). Pulse duration is the most influential variable for both elements (100%), followed by target thickness and laser power. Carbon ionization shows strong sensitivity to main pulse–dominated parameters (target thickness, laser power, beam spot), whereas oxygen exhibits comparatively higher sensitivity to preplasma-related parameters (prepulse-main pulse interval and contrast). In particular, the prepulse–main-pulse interval becomes the third most important factor for oxygen, while contrast also plays a more prominent role for oxygen than for carbon when compared to wavelength and polarization.

In contrast, the behaviour of oxygen displays a stronger sensitivity to preplasma-related parameters. The prepulse–main pulse interval becomes the third most important variable for oxygen (33.2%), while for carbon it plays a comparatively minor role (39.1% but ranking lower due to stronger competitors). Similarly, contrast is significantly more important for oxygen (7.5%) than either wavelength (3.1%) or polarization (5.5%), whereas for carbon the opposite trend is observed − wavelength (10.4%) and polarization (11.0%) remain more influential than contrast (10.2%). Together, these trends demonstrate that oxygen ionization is more sensitive to the preplasma conditions − density, scale length, early-time heating, and the initial distribution of low-energy preplasma electrons − while carbon responds predominantly to the main pulse parameters, which directly modulate hot-electron temperature and sheath strength.

This divergence arises from the underlying atomic physics of the two species. Carbon charge states lie in a lower energy range, so their production is governed mainly by the hot-electron population generated during the main ultrashort pulse [50,51]. Oxygen, however, requires significantly higher ionization thresholds for reaching $O^{5+} - O^{6+}$, making its charge-state distribution much more sensitive to the shape and pre-conditioning of the preplasma before the arrival of the main pulse. Even moderate preplasma formation can modify the effective electron temperature, and introduce additional low- and moderate-energy electron populations that facilitate or suppress particular oxygen charge states [52,53]. Consequently, oxygen responds more strongly to prepulse timing and contrast, while carbon remains primarily governed by the main pulse effects. This distinction between preplasma-dominated and main pulse-dominated behaviour is fully consistent with their respective ionization potentials and the multi-peak quasi-Maxwellian nature of the electron distribution generated under TNSA conditions.

### 6. Confusion-matrices analysis

The performance of the CART classifier for predicting carbon and oxygen ionization states is summarized by the normalized confusion matrices for both the training and test datasets (Fig. 6, Fig. 7). In all cases, the matrices exhibit strong diagonal dominance, demonstrating that the majority of laser shots are correctly classified into their respective charge states. Overall accuracies remain high across datasets, reaching 93% (training – Fig. 6(a) −) and 84% (test – Fig. 6(b) −) for carbon, and 83% (training – Fig. 7(a) −) and 79% (test – Fig. 7(b) −) for oxygen. The moderate reduction in accuracy from training to test data reflects realistic generalization behavior rather than overfitting, indicating that the decision rules extracted by the CART model capture robust, physically meaningful` trends.

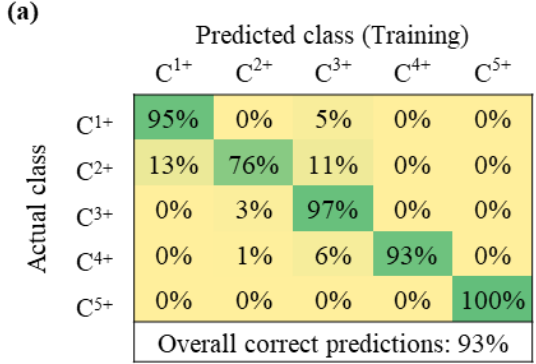
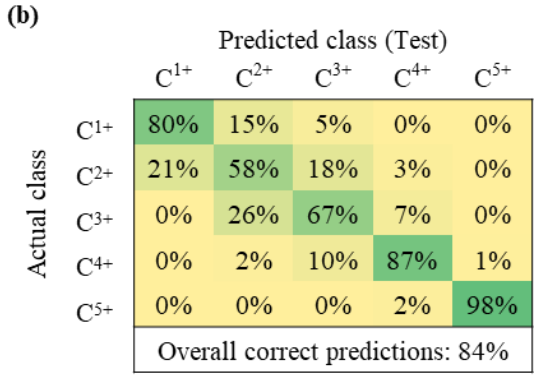

FIG 6. Confusion-matrix heatmaps for CART classification of carbon ionization states: Normalized confusion matrices for the training (**a**) and test (**b**) datasets are shown for carbon charge states ($C^{1+} - C^{5+}$). Each cell represents the percentage of samples of a given true class (rows) that are assigned to a predicted class (columns). Color intensity ranges from green (high percentage, correct classification) to yellow (lower percentage), highlighting classification confidence. Strong diagonal dominance indicates high predictive accuracy, with overall correct classification rates of 93% for the training set and 84% for the test set. Off-diagonal entries primarily correspond to misclassifications between adjacent charge states, reflecting physically expected overlaps in ionization thresholds rather than model instability.

For carbon ions (Fig. 6), classification accuracy is particularly high for the dominant charge state $C^{4+}$, with correct classification rates exceeding 90% in the training set and remaining above ~ 85% in the test set. The highest charge state observed, $C^{5+}$, is identified with near-perfect accuracy in both datasets, reflecting its relatively distinct plasma conditions and reduced overlap with lower-energy electron populations. Misclassifications are largely confined to neighboring charge states, most notably between $C^{2+}$ and $C^{3+}$ or between $C^{3+}$ and $C^{4+}$. This behavior is physically expected, as the ionization energy spacing between successive carbon charge states increases gradually, and modest fluctuations in electron temperature or density can shift the dominant ionization outcome by one charge unit. Importantly, no significant long-range misclassifications (e.g., $C^{1+}$ misidentified as $C^{4+}$) are observed, confirming that the model respects the sequential nature of the ionization ladder.

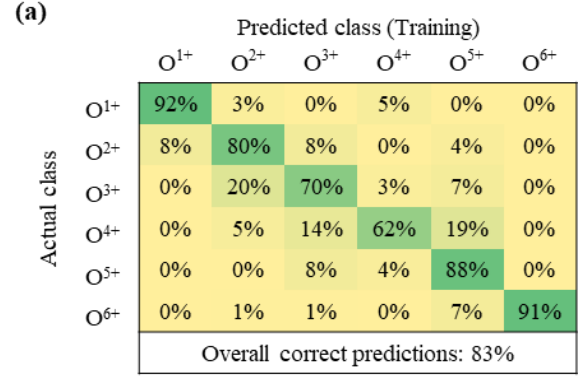
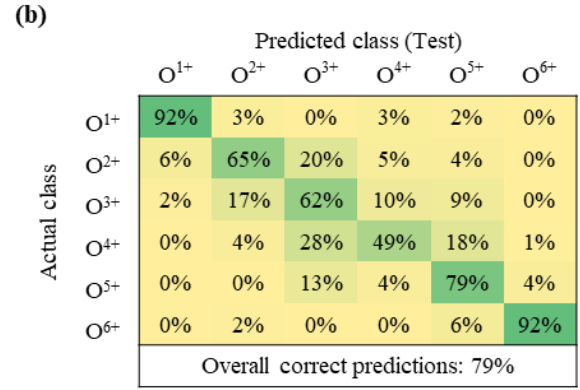

FIG 7. Confusion-matrix heatmaps for CART classification of oxygen ionization states: Normalized confusion matrices for the training (**a**) and test (**b**) datasets are shown for oxygen charge states ($O^{1+} - O^{6+}$). Each cell represents the percentage of samples belonging to a given true class (rows) that are assigned to a predicted class (columns). The color scale transitions from green (strong classification confidence) to yellow (lower percentage). The matrices exhibit clear diagonal dominance, yielding overall correct classification rates of 83% for the training set and 79% for the test set. Misclassifications are predominantly confined to neighboring charge states (e.g., $O^{3+} - O^{4+} - O^{5+}$), consistent with the gradual ionization thresholds and overlapping electron-energy regimes characteristic of TNSA-driven plasmas. The close agreement between training and test performance indicates robust generalization and minimal overfitting of the CART model.

For oxygen ions (Fig. 7), the classification problem is inherently more complex due to the broader distribution of observed charge states and the presence of a dominant high-charge state ($O^{6+}$). Despite this increased complexity, the classifier maintains strong performance, particularly for $O^{6+}$, which is predicted with > 90% accuracy in both training and test sets. This reflects the fact that $O^{6+}$ formation is associated with a relatively well-defined electron energy regime, which is frequently and consistently achieved under the TNSA conditions considered here. Lower and intermediate oxygen charge states ($O^{2+} - O^{4+}$) exhibit slightly higher misclassification rates, primarily with adjacent states. For example, $O^{3+}$ is occasionally misclassified as $O^{2+}$ or $O^{4+}$, and $O^{4+}$ shows partial overlap with $O^{3+}$ and $O^{5+}$. These patterns again reflect the continuous nature of the underlying electron energy distribution and the threshold-driven character of collisional ionization.

A key observation across both elements is that misclassification errors are overwhelmingly local, occurring almost exclusively between adjacent ionization states. This confirms that the CART model is not producing arbitrary decision boundaries but instead learns physically consistent transitions between charge states. Such behavior is characteristic of plasmas where electron energy distributions are well described by multi-peak quasi-Maxwellian populations, leading to partial overlap in the ionization probabilities of neighboring states [49]. Small variations in laser parameters or preplasma conditions can therefore shift the balance between two adjacent ionization stages without fundamentally altering the overall acceleration regime.

The comparison between training and test confusion matrices further demonstrates the robust generalization of the model. While test set accuracies are modestly lower, the overall structure of the matrices remains unchanged, with preserved diagonal dominance and similar off-diagonal patterns. This consistency indicates that the CART classifier does not rely on dataset-specific fluctuations but instead captures stable relationships between the laser and target parameters and ionization outcomes. The absence of significant degradation in test performance supports the conclusion that the model is not overfitted and that the extracted trends are transferable across a wide parameter space.

Finally, the confusion-matrix analysis provides valuable physical insight beyond simple accuracy metrics. The observed classification behavior reinforces the interpretation that ionization in TNSA is governed by smooth, threshold-dependent processes rather than abrupt regime changes. The ability of the CART model to reproduce this behavior confirms its suitability for modeling discrete plasma outputs and supports its use as a predictive tool for ion charge-state distributions in laser-driven ion acceleration experiments.

### 7. Contour and box plot analysis

Because full CART decision trees are large and visually complex, the model outputs describing the ionization states of carbon and oxygen are instead presented using contour and box probability maps. For each element, pulse duration − identified as the most influential parameter for both carbon and oxygen (Fig. 5) − is paired with each remaining continuous predictor to generate two-dimensional maps of the most probable ionization state. Five such contour plots are produced per species, corresponding to pulse duration combined with target thickness, laser power, laser beam spot, prepulse–main pulse interval, and contrast.

For carbon (Fig. 8(a−e)), the resulting maps partition the parameter space into four dominant regions corresponding to the most probable charge-state transitions ($C^{1+} - C^{2+}$, $C^{2+} - C^{3+}$, $C^{3+} - C^{4+}$, and $C^{4+} - C^{5+}$). For oxygen (Fig. 8(f−j)), five analogous regions are observed ($O^{1+} - O^{2+}$ through $O^{5+} - O^{6+}$), reflecting its broader ionization ladder. These regions represent the charge states with the highest probability within each local region of parameter space.

A clear structural trend emerges across all contour plots. When pulse duration is combined with a second parameter of high relative importance (e.g., target thickness), the boundaries between ionization state regions are strongly non-vertical. This indicates that both parameters jointly govern the ionization outcome, consistent with coupled physical effects such as electron heating, confinement time, sheath evolution, and plasma expansion. In contrast, when pulse duration is paired with a parameter of lower relative importance (e.g., contrast), the contour boundaries become nearly vertical. In these cases, the predicted ionization state depends primarily on pulse duration, with the second parameter exerting only a weak influence. This visual behavior directly reflects the quantitative variable-importance ranking derived from the CART analysis and provides an intuitive physical interpretation of parameter coupling.

Discrete parameters − polarization and laser wavelength − are treated separately. For each species (Fig. 9), two diagrams are constructed per discrete variable: one for linear and one for circular polarization, and one for 815 nm and one for 1054 nm wavelength. These plots display pulse duration on the horizontal axis and ionization state on the vertical axis, with shaded boxes indicating the most probable charge-states transition. The similarity of the box patterns across polarization and wavelength confirms their limited influence on ionization under classical TNSA conditions, while still allowing subtle shifts at specific pulse durations.

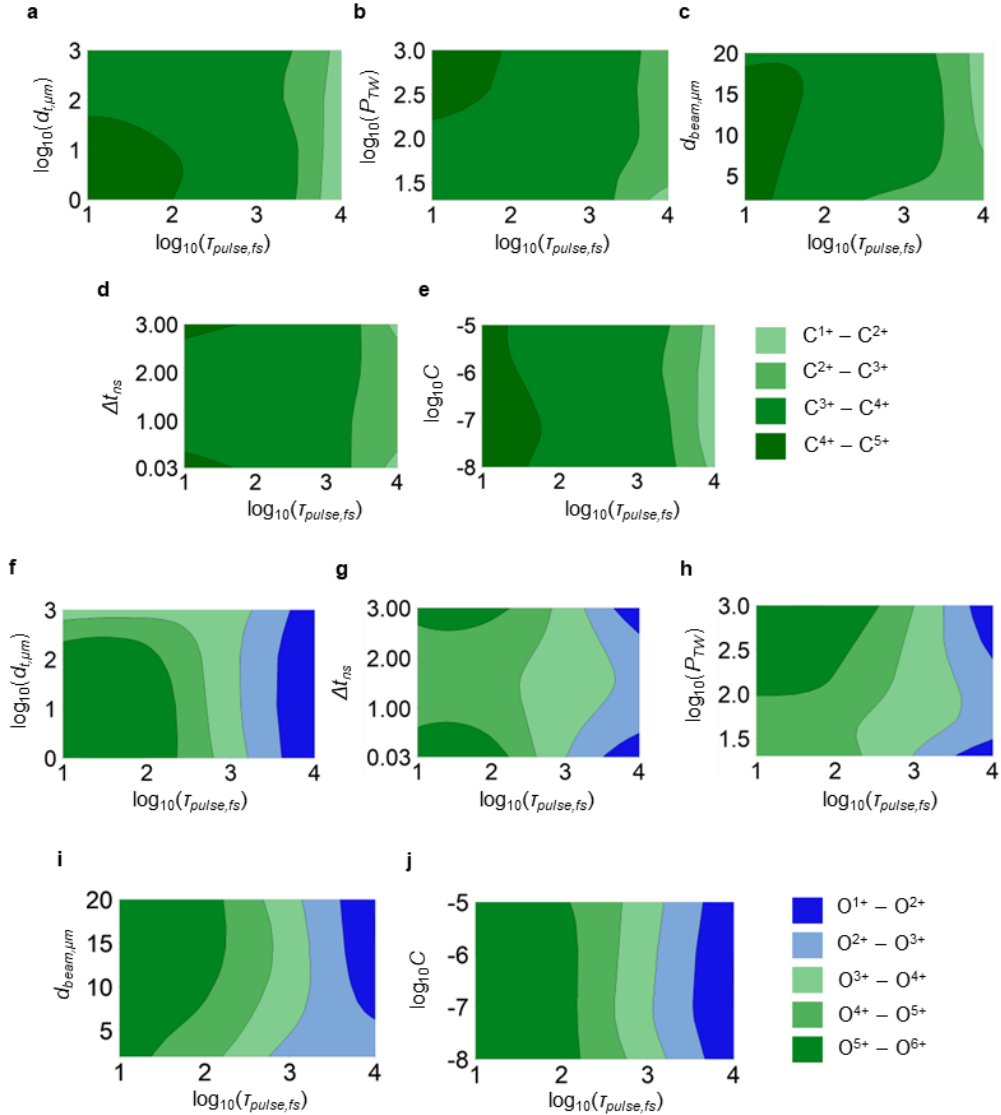

FIG 8. Contour maps of the most probable ionization states of carbon and oxygen as a function of pulse duration and continuous parameters: Each panel shows the dominant charge-state transition predicted by the CART model as a function of pulse duration and one additional continuous input parameter. Separate contour maps are presented for carbon and oxygen. For each species, the panels are ordered from highest to lowest relative importance of the second parameter: (**a**, **f**) target thickness, (**b**, **h**) laser power, (**c**, **i**) laser beam spot, (**d**, **g**) prepulse–main pulse interval, and (**e**, **j**) contrast. Color-coded regions indicate the most probable ionization-state ranges, spanning $C^{1+} - C^{2+}$ through $C^{4+} - C^{5+}$ for carbon and $O^{1+} - O^{2+}$ through $O^{5+} - O^{6+}$ for oxygen, within each local region of the parameter space. Strongly non-vertical boundaries indicate coupled dependence on both pulse duration and the second parameter, whereas nearly vertical boundaries reflect dominance of pulse duration. These maps visualize how carbon and oxygen ionization states emerge from the combined influence of laser and target parameters under TNSA-relevant conditions.

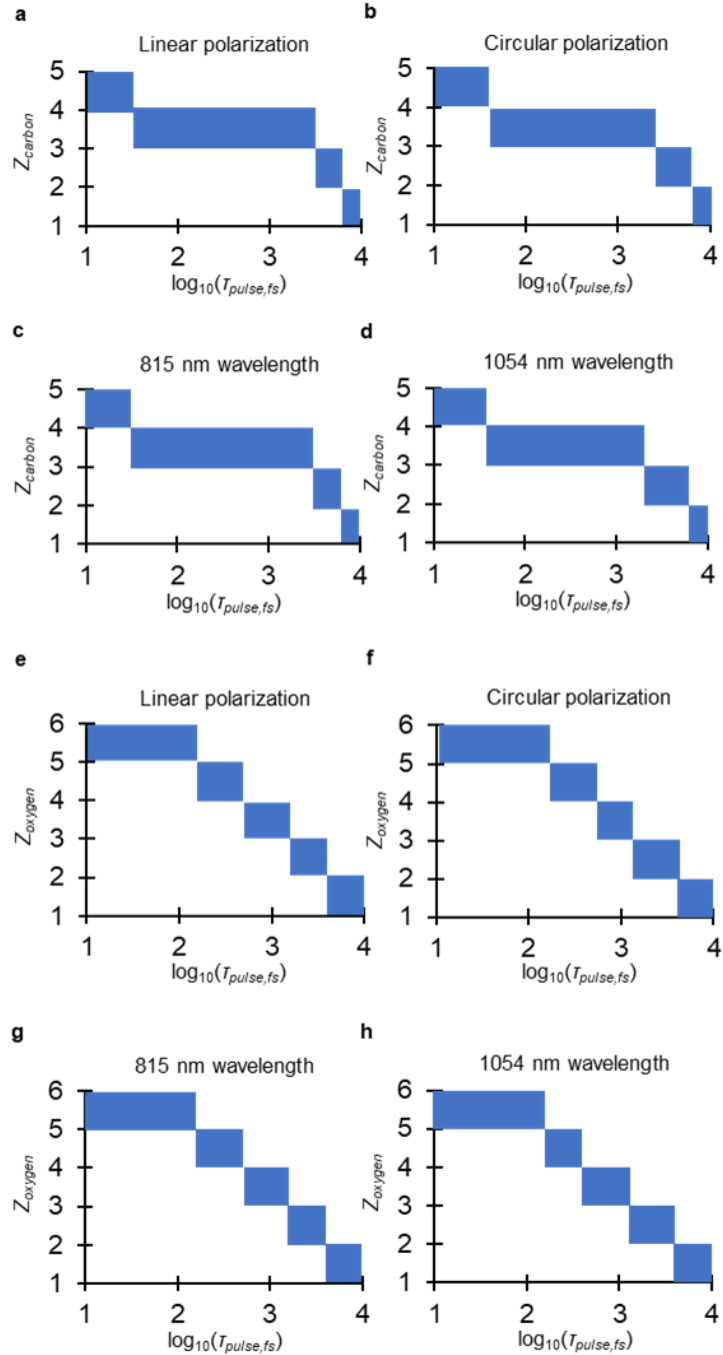

FIG 9. Box-plot probability maps of dominant carbon and oxygen ionization states for discrete laser parameters: Box plots summarize the dominant charge-state transitions predicted by the CART model as a function of pulse duration for discrete input parameters. For each species, separate panels are shown for (**a**, **e**) linear polarization, (**b**, **f**) circular polarization, (**c**, **g**) laser wavelength of 815 nm, and (**d**, **h**) laser wavelength of 1054 nm, with carbon and oxygen presented in the upper and lower rows, respectively. The horizontal axis corresponds to pulse duration, while the vertical axis denotes the ionization state. Shaded boxes indicate the most probable charge-state transition for different pulse durations. The overall similarity of the box patterns across polarization and wavelength confirms the limited influence of these discrete parameters under classical TNSA conditions, while subtle shifts at specific pulse durations highlight secondary modulation effects.

Beyond interpretation, these contour and box plots provide a practical predictive tool. To estimate the most probable ionization state for a given experimental configuration, a researcher can follow the methodology below:

1. Identify the pulse duration of interest.

2. Locate the corresponding values of the remaining relevant parameters in the appropriate contour and box plots.

3. Read off the dominant ionization-state region.

Because ionization is inherently probabilistic and governed by overlapping electron-energy distributions, it is possible that different contour and box plots suggest adjacent charge-state ranges for the same parameter set. In such cases, the most probable ionization state is identified as the overlapping charge state common to the predicted ranges. For example, if some plots indicate $C^{3+} - C^{4+}$ while others indicate $C^{4+} - C^{5+}$, $C^{4+}$ is selected as the statistically dominant and physically consistent prediction. This selection criterion reflects the probabilistic nature of ionization and corresponds to the charge state with the highest joint likelihood across the multidimensional parameter space.

Overall, the contour and box plot analysis synthesizes the statistical CART results into an interpretable, physics-consistent framework. It demonstrates how ionization states emerge from the coupled interplay of pulse duration with other laser and target parameters and provides a transferable methodology for predicting charge-state distributions across a wide range of TNSA-relevant conditions. Together with the variable importance analysis and confusion matrix validation, these results form a comprehensive statistical and physical characterization of ion charge-state production in the TNSA regime.

## IV. CONCLUSIONS

In this work, we develop and apply statistically rigorous, physics-informed methods to quantify how laser and target parameters jointly govern proton and ion acceleration and charge-state formation in the Target Normal Sheath Acceleration (TNSA) regime. Using a unified multiphysics model, validated against experiments from multiple laser systems and achieving predictive accuracy exceeding 95%, we derive multivariate scaling laws and probability maps that correlate the cutoff energies, beam divergences, and ionization states of proton, carbon, and oxygen ion beams with a broad parameter space relevant to contemporary short- and ultrashort-pulse laser systems. This parameter space includes pulse duration, target thickness, laser power, laser beam spot, prepulse–main pulse interval, polarization, laser wavelength, and contrast.

Several key physical observations emerge from this analysis. First, for continuous beam properties − cutoff energies and divergence − all laser and target parameters except polarization contribute significantly and are therefore retained in the multivariate scaling laws. The resulting models achieve robust predictive performance, with $R^2$ values of approximately 70% for proton and ion cutoff energies and exceeding 98% for beam divergences, while standard, adjusted, and 10-fold cross-validated $R^2$ remain in close agreement. This consistency indicates minimal overfitting and confirms that the scaling laws capture genuine physical dependencies rather than dataset-specific correlations.

Second, the comparatively lower $R^2$ values for ion cutoff energies (≈ 70%) are fully consistent with the underlying physics and the relevant literature. Cutoff energies are influenced by stochastic micro-scale plasma dynamics, sheath-field fluctuations, and instabilities, which are not entirely captured by the macroscopic laser and target parameters included in the scaling laws. In contrast, beam divergence exhibits much higher predictability ($R^2 > 98\%$) because it is largely determined by more deterministic geometric and temporal properties of the accelerating sheath.

Third, to investigate carbon and oxygen ionization states, a CART model is implemented, which captures both the statistical structure and the underlying physics of charge-state formation in the TNSA regime. Carbon exhibits a strongly peaked distribution dominated by $C^{4+}$, while oxygen displays a broader distribution with $O^{6+}$ as the most probable state. These outcomes are physically consistent with the ionization potentials of the two species and with the multi-peak quasi-Maxwellian electron energy distributions characteristic of TNSA-driven plasmas. In both cases, dominant charge states emerge from the overlap between accessible electron-energy populations and the corresponding ionization thresholds, rather than from extreme high-energy tails.

Fourth, the relative variable importance analysis further reveals a clear species-dependent sensitivity to laser and target parameters. While pulse duration is the dominant driver for both carbon and oxygen, carbon ionization is primarily governed by main-pulse parameters such as target thickness and laser power, whereas oxygen ionization shows enhanced sensitivity to preplasma-related parameters, including prepulse–main pulse interval and contrast. This distinction reflects the higher ionization thresholds required for oxygen and highlights the critical role of preplasma formation in shaping its charge-state distribution.

Fifth, the CART classifier demonstrates robust predictive performance and generalization, with misclassifications confined almost exclusively to adjacent charge states. This behavior confirms that the model respects the sequential and threshold-driven nature of collisional ionization in laser-produced plasmas.

Overall, the results show that TNSA ion acceleration is intrinsically multivariate and cannot be reliably described by reduced-parameter models. By integrating validated multiphysics-based regression and classification, this work provides a predictive and interpretable framework for optimizing laser-driven ion sources. The scaling laws and probability contour maps and box plots presented here offer practical guidance for experimental design and establish a foundation for extending multivariate statistical modeling to more advanced acceleration regimes.

## ACKNOWLEDGMENTS
The author acknowledges that no external funding was received for this work.

## COMPETING INTERESTS
The author has not any conflicts of interest to disclose concerning this study.

## DATA AVAILABILITY
All data that support the findings of this study are available from the corresponding author upon reasonable request.

## CODE AVAILABILITY
All the codes developed in this study are available from the corresponding author upon reasonable request.